\documentclass[A4,11pt,epsf,oneside]{article}
\usepackage{amssymb, amscd, amsmath,amsthm,amsfonts}
\usepackage{epsfig}
\usepackage{xcolor}
\usepackage{capt-of}
\usepackage{float}
\restylefloat{table}

\begin{document}
\renewcommand{\thefootnote} {\fnsymbol{footnote}}
\setcounter{page}{1}

\title{\textbf{Complete classification of spherically symmetric static spacetimes via Noether symmetries}}
\author{\textbf{Farhad Ali, Tooba Feroze and Sajid Ali }
 \vspace{1cm}\\
{\normalsize \it \textbf{School of Natural Sciences,
}}\\
{\normalsize \it \textbf{National University of Sciences and Technology, H-12
Islamabad, Pakistan.}}}

       \baselineskip=5mm
\baselineskip=7mm
 \maketitle
\begin{abstract}
In this paper we give a complete classification of spherically symmetric static space-times by their Noether symmetries. The determining equations for Noether symmetries are obtained by using the usual Lagrangian of a general spherically symmetric static spacetime which are integrated for each case. In particular we observe that spherically symmetric static spacetimes are categorized into six distinct classes corresponding to Noether algebra of dimensions 5, 6, 7, 9, 11 and 17. Using Noether`s theorem we also write down the first integrals for each class of such spacetimes corresponding to their Noether symmetries.

\end{abstract}

\section{Introduction}
Spherical symmetric static solutions of Einstein's field equations are of great importance
in general relativity for they possess the properties of being static and asymptotically flat
which was proved by Birkhoff \cite{1}. For a spherical symmetric spacetime there are exactly
three rotational Killing vector fields that preserve the metric forming $SO(3)$ as the isometry group
of symmetries of these spacetimes. The study of spherically symmetric spacetimes is interesting as it helps in giving the understanding of phenomena of gravitational collapse and black holes, a widely known subject in the literature. For example the
Schwarzschild solution is a non-trivial exact solution of the Einstein's field equations which is spherically symmetric that describes the gravitational
field exterior to a static, spherical, uncharged mass without angular momentum and isolated from all other mass.

The search for spherically symmetric spacetimes is an important task and due to their significance in understanding

 the dynamics around black-holes it is crucial to classify them with respect to their physical properties. Hence it
  would be interesting to find out the general form of these spacetimes along with a detailed characterization of the
   first integrals of corresponding geodesic equations. Besides in order to obtain those quantities which remain invariant
    under the geodesic motions yield significant physical information. In \cite{2}, Isabel Cordero-Carrión provided a procedure
     based on geometric arguments to obtain maximal foliations of spherically symmetric spacetimes which they later use to calibrate
      numerical relativity complex codes. The classification of these spacetimes based on their local conformal symmetries is done in
       \cite{10}. On the other hand classification of plane-, cylindrically- and spherically-symmetric spacetimes with respect to their
        Killing vectors, homotheties, Ricci collineations, curvature collineations have been done in \cite{3, 15}.

In a recent paper, Farhad Ali and Tooba Feroze \cite{4} used the symmetry method approach to find the complete classification of
plane-symmetric spacetimes from Noether symmetries and also present the first integral in each case. In this paper we employ
 symmetry methods to provide complete classification of spherically symmetric static spacetimes with respect to the Noether
 symmetries they possess. Later, we use famous Noether's theorem to write down the first integrals of each spacetime.
 In this way we are not only be able to recover all the known cases but also list the
corresponding first integrals of solutions of EFEs, which are given
in standard gravitational units, $c = G = 1$
as,\begin{equation}R_{\mu\nu}-\frac{1}{2}Rg_{\mu\nu}=8\kappa
T_{\mu\nu}.
\end{equation}

The general form of a spherically symmetric static space-time is
\begin{equation}
ds^2=e^{\nu(r)}dt^2-e^{\mu(r)}dr^2-e^{\lambda(r)}\mbox{d}\Omega^{2}.
\end{equation}
where $\mbox{d}\Omega^2 = \mbox{d}\theta^2 + \sin^{2}\theta
\mbox{d}\phi^2,$  and both $\nu$ and $\mu$ are arbitrary functions
of radial coordinate `$r$'. It is seen that $e^{\lambda(r)}$ can be
one of the two forms (i) $\beta^2$ or (ii) $r^2$, where $\beta$ is
some constant \cite{5}, we absorbed $\beta^2$ in the definition of
$\mbox{d}\Omega^{2}$ . We write down the determining equations using
the corresponding Lagrangian of
 above spacetime and study the complete integrability of those equations for each case.

The plan of the paper is as follows. In the section two we describe
basic definitions and structure of Noether symmetries along. In
section three we write down the determining equations for
spherically symmetric static spacetimes which is a system of
nineteen linear partial differential equations (PDEs). We obtain
several cases for different forms of `$\nu$' and `$\mu$' while
integrating the PDEs that give rise to a complete classification of
spherically symmetric static spacetimes with Noether algebra of
dimensions 5, 6, 7, 9, 11 and 17. The characterization of first
integrals along geodesic motions is also carried out in the same
section. The conclusion is given in the third section.

\section{Preliminaries}
It is well-known that a general spherically symmetric static spacetime admits usual Lagrangian $L= L(t,r,\theta,\phi)$
\begin{equation}
L=e^{\nu(r)}\dot{t}^2-e^{\mu(r)}\dot{r}^2-e^{\lambda{(r)}} (\dot{\theta}^2+\sin^{2}\theta\dot{\phi}^2),
\end{equation}
where
$``."$ denotes differentiation with respect to arc-length parameter `$s$'. A symmetry $\mathbf{X}$ of
the Lagrangian $L$ that leaves the action of a spherically symmetric static spacetime invariant is a Noether symmetry if it satisfies the
following equation,
\begin{equation}
\mathbf X^{(1)}L+ \mathbf{D}(\xi)L = \mathbf{D} A,
\end{equation}
where
\begin{equation}
\mathbf X^{(1)}=\mathbf X+\eta^0_{,s}\frac{\partial}{\partial
t}+\eta^1_{,s}\frac{\partial}{\partial
r}+\eta^2_{,s}\frac{\partial}{\partial
\theta}+\eta^3_{,s}\frac{\partial}{\partial \phi},
\end{equation}
is the first order prolongation of
\begin{equation}
\mathbf
X=\xi\frac{\partial}{\partial s}+\eta^0\frac{\partial}{\partial
t}+\eta^1\frac{\partial}{\partial r}+\eta^2\frac{\partial}{\partial
\theta}+\eta^3\frac{\partial}{\partial \phi},
\end{equation} $\mathbf{D}$ is the standard total derivative operator given by
\begin{equation}
\mathbf{D}=\frac{\partial}{\partial s}+\dot{t}\frac{\partial}{\partial
t}+\dot{r}\frac{\partial}{\partial
r}+\dot{\theta}\frac{\partial}{\partial
\theta}+\dot{\phi}\frac{\partial}{\partial\phi},
\end{equation}
and $ A$ is some gauge function. The coefficients of Noether
symmetry, namely, $\xi$, $\eta^i$  and gauge function $A$ are
functions of $(s,t,r,\theta,\phi)$. The coefficients of prolonged
operator $\mathbf{X}^{(1)}$, namely, $\eta^i_{,s}$, are functions of
$(s,t,r,\theta,\phi,\dot{t},\dot{r},\dot{\theta},\dot{\phi})$ and
are determined by
\begin{equation}
\eta^{i}_{,s} = \mathbf{D} (\eta^{i}) - \dot{u}^{i}\mathbf{D}(\xi ),
\end{equation}
where $u^{i}$ refers to the space of dependent variables $(t,r,\theta,\phi)$ for $i=0,1,2,3$. Using the same identification
we state famous Noether's theorem.
\newline
\textbf{Noether Theorem} \newline
If $\mathbf{X}$ is a Noether symmetry of a given Lagrangian $L$ with respect to the gauge function $A$, then the quantity
\begin{equation}
I = A - \left ( \xi L + \left ( \eta^{i} - \xi \dot{u}^{i}\right )
\frac{\partial L}{\partial \dot{u}^{i}}
 \right),
\end{equation}
is annihilated by the total derivative operator, i.e., $\mathbf{D}I=0$.

\section{Classification Results and Computational Remarks}
By substituting the value of Lagrangian (3) in equation (4) and
comparing the coefficients of all monomials we obtain the following
system of nineteen linear partial differential equations (PDEs)
\begin{equation}\begin{split}
 &
\xi_t=0 , \quad \xi_r=0 ,\quad\xi_\theta=0 ,\quad \xi_\phi=0 ,\\
&
A_s=0,\quad A_{t}-2e^{\nu(r)}\eta^{0}_s=0 ,\quad A_{r}+2\eta^{1}_s=0,\\
&
A_{\theta}+2e^{\mu(x)}\eta^2_{s}
 =0 ,\quad A_{\phi}+2e^{\mu(r)}\eta^3_{s}=0,\\
&
\xi_s - \mu_r(r)\eta^1 - 2\eta^1_r =0, \quad \xi_s-\nu_r(r)\eta^1-2\eta^0_t=0
\\
&
\xi_s-\frac{2}{r}\eta^1 - 2\eta^2_\theta =0, \quad
\xi_s-\frac{2}{r}\eta^1 - 2\cot\theta\eta^2 -2\eta^3_\phi=0,
\\
&
\eta^2_\phi+\sin^2\theta\eta^3_\theta=0 ,\quad
e^{\nu(r)}\eta^0_r-e^{\mu(x)}\eta^1_t=0
,\\
&
e^{\mu(r)}\eta^1_\theta + r^2\eta^2_r = 0
,\quad e^{\nu(r)}\eta^0_\theta-r^2\eta^2_t=0,
\quad
\\
& e^{\nu(r)}\eta^0_\phi-r^2\sin^2\theta\eta^3_t=0 , \quad
e^{\mu(r)}\eta^1_\phi + r^2\sin^2\theta\eta^3_r=0 .
\end{split}\end{equation} We intend to classify all spherically symmetric
static spacetimes with respect to their Noether symmetries by
finding the solutions of above system of PDEs. We have used Computer
Algebra System (CAS) Maple$-$17 to carry out the case-splitting with
the help of an important algorithm `rifsimp' which is essentially an
extension of the Gaussian elimination and Groebner basis algorithms
that is used to simplify overdetermined systems of polynomially
nonlinear PDEs or ODEs and inequalities and bring them into a useful
form. In the following section we enlist spherically symmetric
static spacetimes, their Noether symmetries and relative first
integrals. We also present the Noether algebra of Noether symmetries
in the cases that are  not known in the literature.

In order to solve system of PDEs $(10)$, we note that the first equation simply implies that $\xi$ can only be
 a function of arc-length parameter, i.e., $\xi(s)$. To keep a distinction between Killing vector fields and Noether
  symmetries we use a different letter, namely, $\mathbf{Y}$ for those Noether symmetries which are not Killing vector fields.
   It is also remarked that a static spacetime always admit a time-like Killing vector field. Moreover, the Lagrangian $(3)$
   does not depend upon `$t$' explicitly therefore the time-like Killing vector field appears as a Noether symmetry in each case.
    Furthermore, the Lagrangian $(3)$ is spherically symmetric, therefore, the Lie algebra of Killing vector fields $so(3)$
    corresponding to the Lie group $SO(3)$ is intrinsically admitted by each
    spacetime. It is also important to mention here that a static spacetime
  always admit a time-like Killing vector field. Hence,
\begin{align}
\mathbf{X_0}=\frac{\partial}{\partial t}, \quad
\mathbf{Y_0}=\frac{\partial}{\partial s},
\end{align}
\begin{equation}
 \mathbf{X_1}=\frac{\partial}{\partial
\phi}, \quad
\mathbf{X_2}=\cos\phi\frac{\partial}{\partial\theta}-\cot\theta\sin\phi\frac{\partial}{\partial\phi},
\quad
\mathbf{X_3}=\sin\phi\frac{\partial}{\partial\theta}+\cot\theta\cos\phi\frac{\partial}{\partial\phi}\,.
\end{equation}
form the basis of minimal $5-$dimensional Noether algebra, in which
$\mathbf{Y}_{0}$ is not a Killing vector field of spherically
symmetric spacetime (2). The Noether algebra of these five Noether
symmetries is, $[\mathbf{X_1},\mathbf{X_3}]=\mathbf{X_2}$,
$[\mathbf{X_2},\mathbf{X_3}]=-\mathbf{X_1}$,
 $[\mathbf{X_i},\mathbf{X_j}]=0$ and $[\mathbf{X_i},\mathbf{Y_0}]=0$
otherwise, and is identified with the associated group $SO(3) \times    \mathbb{R}^2 $.
\newline \newline
\textbf{I - Spherically Symmetric Static Spacetimes with Five Noether Symmetries}
\newline
Some examples of spacetimes that admit minimal set of Noether symmetries (five
symmetries) appeared during the calculations are given in Table 1.
\\
\renewcommand{\arraystretch}{2}
\begin{table}[H]
\begin{center}
\captionof{table}{Forms of $\mu$ and $\nu$}
\begin{tabular}{|c|c|c|}
 \hline
  No.& $\nu(r)$ & $\mu(r)$\\
  \hline
  1. & $\ln\left(\frac{r}{\alpha}\right )^2$ & arbitrary \\
  \hline
  2. & $\ln\left (1-(\frac{r}{\alpha})^2\right)$ & arbitrary \\
  \hline
  3. & $\ln(\frac{r}{\alpha})^2$ & $-\ln\left (1-(\frac{r}{\alpha})^2\right)$ \\
  \hline
  4. & arbitrary& $-\ln(1-(\frac{r}{\alpha})^2)$  \\
  \hline
  5. & $\ln(1-\frac{\alpha}{r})$& $-\ln(1-\frac{\alpha}{r})$  \\
  \hline
  \end{tabular} \end{center}\end{table}
 The Noether symmetries and corresponding first integrals are listed in the
following Table 2
\begin{table}[H]
\begin{center}
\captionof{table}{First Integrals}
\begin{tabular}{|c||c|}\hline

                Gen & First integrals \\
                \hline
               $ \mathbf{X_0}$ & $\phi_0=e^{\nu(r)}\dot{t} $ \\
               \hline
               $ \mathbf{X_1} $& $\phi_1=r^2\sin^2\theta\dot{\phi}$ \\
               \hline
               $ \mathbf{X_2}$ & $\phi_2=r^2\left (\cos\phi\dot{\theta} -\cot\theta\sin\phi\dot{\phi} \right)$ \\
               \hline
                $\mathbf{X_3}$ & $\phi_3=r^2\left (\sin{\phi}\dot{\theta}+\cot\theta\cos\phi \dot{\phi} \right)$\\
                \hline
               $\mathbf{Y_0}$&$\phi_4=e^{\nu(r)}\dot{t}^2-e^{\mu(r)}\dot{r}^2-r^2 \left(\dot{\theta}^2+\sin^2
               \theta\dot{\phi}^2 \right )$\\
               \hline

\end{tabular}\end{center}\end{table}
with constant value of the gauge function, i.e., $ A=\mbox{constant}$.
\newline \newline
\textbf{II - Spherically Symmetric Static Spacetimes with Six Noether Symmetries}
\newline
There are two distinct classes of spherically symmetric static spacetimes admitting six Noether symmetries. In particular
we get \\
\begin{equation}
(1): \hspace{2cm}
\mbox{d}s^2=\left (\frac{r}{a} \right )^\alpha
\mbox{d}t^2-\mbox{d}r^2-r^2 \mbox{d}\Omega^{2} ,\quad \alpha \neq 0,2
\end{equation}
which apart from minimal five-dimensional Noether algebra aslo admit an additional Noether symmetry corresponding to
the scaling transformation $(s,t,r) \longrightarrow (\lambda s, \lambda ^{p} t, \lambda^{1/2}r)$, given by
\begin{equation}
\mathbf{Y_1}=s\frac{\partial}{\partial s}+pt\frac{\partial}{\partial t}
+\frac{r}{2}\frac{\partial}{\partial r},\quad p=\frac{2-\alpha}{4}
\end{equation}
forming
a $6-$dimensional Noether algebra. This induces a scale-invariant spherically symmetric static spacetime. The corresponding first integral is
\begin{equation}
\phi_6 = s \left (\left (\frac{r}{a}\right )^\alpha
\dot{t}^2-\dot{r}^2-r^2(\dot{\theta}^2+\sin^2\theta \dot{
\phi}^2)\right )+\frac{\alpha-2}{2}\left (\frac{r}{a}\right
)^2t\dot{t}+r\dot{r}
\end{equation}
The other spacetime with $6-$dimensional Noether algebra is given by
\begin{equation}
(2):\hspace{2cm}\mbox{d}s^2=\mbox{d}t^2-e^{\mu(r)}\mbox{d}r^2-r^2\mbox{d}\Omega^{2},
\quad \mu(r)\neq \ln \left (1-\frac{r^2}{b^2}\right)^{-1},\mu(r)\neq
constant
\end{equation}
with an additional Noether symmetry relative to a non-trivial guage term
\begin{equation}
\mathbf{Y_1}=s\frac{\partial}{\partial
t}, \quad A=2t.
\end{equation}
The first integral corresponding to
$\mathbf{Y_1}$ is      $\phi_6=t-s\dot{t}$.
\newline \newline
\textbf{III - Spherically Symmetric Static Spacetimes with Seven Noether Symmetries}
\newline
There arise five spacetimes with seven Noether symmetries in which
four cases contains the group of six Killing vector fields whereas
one case contain only the minimal group of Killing vectors while the
other two symmetries are Noether symmetries. We discuss them
separately.

The three metrics are given by
\begin{align*}
&
(1): \hspace{1cm}\mbox{d}s^2=e^{r/b}\mbox{d}t^2-\mbox{d}r^2-\mbox{d}\Omega^{2}, \quad b\neq 0, \quad \beta \neq 0 \\
&
 (2): \hspace{1cm}\mbox{d}s^2=\sec^2\left(\frac{r}{a}\right)\mbox{d}t^2-\sec^2\left(\frac{r}{a}\right)\mbox{d}r^2-\mbox{d}\Omega^{2}, \quad a\neq 0 \\
 &
 (3): \hspace{1cm}
 \mbox{d}s^2=\left (1-\frac{r^2}{b^2}\right)\mbox{d}t^2-\left (1-\frac{r^2}{b^2}\right )^{-1}\mbox{d}r^2-\mbox{d}\Omega^{2}, \quad b\neq
 0\\&
 (4): \hspace{1cm}
 \mbox{d}s^2=\frac{\alpha^2}{r^2}\mbox{d}t^2-\frac{\alpha^2}{r^2}\mbox{d}r^2-\mbox{d}\Omega^{2},\quad
 \alpha\neq0,\quad
\end{align*}
have two additional symmetries along with the
minimal set of symmetries
\begin{align*}
&
\mathbf{X_{4,1}}=t\frac{\partial}{\partial r}-b\left (e^{-r/b}+\frac{t^2}{4b^2} \right )\frac{\partial}{\partial t}\quad , \\
&
\mathbf{X_{5,1}}=\frac{\partial}{\partial r}-\frac{t}{2b}\frac{\partial}{\partial
t}\quad ,\\
&
\mathbf{X_{4,2}}=\sin\left(\frac{r}{a}\right)\cos\left(\frac{t}{a}\right)\frac{\partial}{\partial
t}+\sin\left(\frac{t}{a}\right)\cos\left(\frac{r}{a}\right)\frac{\partial}{\partial
r}\quad , \\
&
\mathbf{X_{5,2}}=\cos\left(\frac{t}{a}\right)\cos\left(\frac{r}{a}\right)\frac{\partial}{\partial
r}-\sin\left(\frac{r}{a}\right)\sin\left(\frac{t}{a}\right)\frac{\partial}{\partial
t}\quad ,
\\
&
\mathbf{X_{4,3}}=-\frac{rbe^{t/b}}{\sqrt{r^2-b^2}}\frac{\partial}{\partial t}+\sqrt{r^2-b^2}e^{t/b}\frac{\partial}{\partial r}\quad ,\\
&
 \mathbf{X_{5,3}}=\frac{rbe^{-t/b}}{\sqrt{r^2-b^2}}\frac{\partial}{\partial t}+\sqrt{r^2-b^2}e^{-t/b}\frac{\partial}{\partial
 r}\quad\\&
 \mathbf{X_{4,4}}=\frac{t^2+r^2}{2}\frac{\partial}{\partial t}+rt\frac{\partial}{\partial
 r}\\&
 \mathbf{X_{5,4}}=t\frac{\partial}{\partial t}+r\frac{\partial}{\partial r}
\end{align*}
respectively, where the new subscript refers to the case distinction and same will be followed from here after.
 The corresponding first integrals are mentioned in the Table 3. The Lie
 algebra of the symmetries of metric (2) is\begin{align*}&[\mathbf{X_1},\mathbf{X_3}]=\mathbf{X_2},\quad  [\mathbf{X_2},\mathbf{X_3}]=-\mathbf{X_1}
 ,\quad,[\mathbf{X}_0,\mathbf{X_{4,2}}]=\frac{1}{a}\mathbf{X_{5,2}},\\&
[\mathbf{X_0},\mathbf{X_{5,2}}]=-\frac{1}{a}\mathbf{X_{4,2}},\quad
 [\mathbf{X_i},\mathbf{X_j}]=0,\quad [\mathbf{X_i},\mathbf{Y_0}]=0,\qquad otherwise\end{align*}
and the Lie algebra of the symmetries of metric (4) is
\begin{align*}&[\mathbf{X_1},\mathbf{X_3}]=\mathbf{X_2},\quad  [\mathbf{X_2},\mathbf{X_3}]=-\mathbf{X_1}
 ,\quad,[\mathbf{X_0},\mathbf{X_{4,4}}]=\mathbf{X_{5,4}},\quad [\mathbf{X_{4,4}},\mathbf{X_{5,4}}]=-\mathbf{X_{4,4}}\\&
[\mathbf{X_0},\mathbf{X_{5,4}}]=\mathbf{X_0},\quad
 [\mathbf{X_i},\mathbf{X_j}]=0,\quad [\mathbf{X_i},\mathbf{Y_0}]=0,\qquad otherwise\end{align*}

  The following spacetime
\begin{equation}(5) \hspace{2cm}
\mbox{d}s^2=\left (\frac{r}{a}\right )^2\mbox{d}t^2-\mbox{d}r^2-r^2\mbox{d}\Omega^{2}
\end{equation}
contain two non-trivial Noether symmetries
\begin{align*}
&
\mathbf{Y_1}=s\frac{\partial}{\partial s}+\frac{r}{2}\frac{\partial}{\partial r}, \\
&
\mathbf{Y_2}=\frac{s^2}{2}\frac{\partial}{\partial
s}+\frac{rs}{2}\frac{\partial}{\partial r}, \quad A=\frac{-r^2}{2},
\end{align*}
whose first integrals are also given in the same Table 3.

\begin{table}[H]
\begin{center}
\captionof{table}{First Integrals of Spacetimes admitting 7 Noether
Symmetries}
\begin{tabular}{|c|c|}\hline

               Gen & First integrals
\\
\hline
$\mathbf{X_{4,1}}$&$\phi_5=b \left (1+\frac{t^2e^{r/b}}{4b^2}\right )\dot{t}+t\dot{r}$\\
\hline
$\mathbf{X_{5,1}}$&$\phi_6=\frac{te^{r/b}}{b}\dot{t}+2\dot{r}$\\
\hline
$\mathbf{X_{4,2}}$&$\phi_5=\sec^2\frac{r}{a}[\dot{t}\sin\frac{r}{a}\cos\frac{t}{a}-\dot{r}\sin\frac{t}{a}\cos\frac{r}{a}]$\\
\hline
$\mathbf{X_{5,2}}$&$\phi_6=-\sec^2\frac{r}{a}[\dot{t}\sin\frac{r}{a}\sin\frac{t}{a}+\dot{r}\cos\frac{t}{a}\cos\frac{r}{a}]$\\
\hline
$\mathbf{X_{4,3}}$&$\phi_5=e^{\frac{t}{b}}[\frac{r\dot{t}\sqrt{r^2-b^2}}{b}+\frac{b^2\dot{r}}{\sqrt{r^2-b^2}}]$\\
\hline
$\mathbf{X_{5,3}}$&$\phi_6=e^{\frac{-t}{b}}[-\frac{r\dot{t}\sqrt{r^2-b^2}}{b}+\frac{b^2\dot{r}}{\sqrt{r^2-b^2}}]$\\
\hline
$\mathbf{X_{4,4}}$&$\phi_5=\frac{(t^2+r^2)\dot{t}}{r^2}+\frac{t\dot{r}}{r}$\\
\hline $\mathbf{X_{5,4}}$&$\phi_6=2[\frac{t\dot{t}}{r^2}-\dot{r}{r}]$\\
\hline
$\mathbf{Y_{1}}$&$\phi_5=sL-r\dot{r}+2s\dot{r}^2$\\
\hline
$\mathbf{Y_{2}}$&$\phi_6=2s^2L+2(s\dot{r}^2-sr\dot{r})+r^2$\\
\hline
\end{tabular}\end{center} \end{table}
\textbf{IV - Spherically Symmetric Static Spacetimes with Nine Noether Symmetries}
\newline
This section contains some well known and important spacetimes. Here,
we have five different cases of spacetimes in which three contain two additional
Noether symmetries and one case contains one extra Noether symmetry besides others
which are all Killing vector fields. We have the following results (spacetimes) with nine Noether symmetries:\\
\begin{align*}
&
(1): \hspace{1cm}ds^2=dt^2-dr^2-\mbox{d}\Omega^{2}, \quad \beta \neq 0 \\
&
(2): \hspace{1cm}ds^2=\frac{\beta^2}{r^2}dt^2-\frac{\beta^4}{r^4}dr^2-\mbox{d}\Omega^{2}, \\
&
(3): \hspace{1cm}ds^2=\left(1+\frac{r}{b}\right)^2dt^2-dr^2-\mbox{d}\Omega^{2} \\
& (4):
\hspace{1cm}ds^2=dt^2-\frac{dr^2}{1-\frac{r^2}{b^2}}-r^2\mbox{d}\Omega^{2}.
\end{align*}
The first three space times correspond to the famous
Bertotti-Robinson like solutions of Einstein's field equations which
describes a universe with uniform magnetic field whereas the last
case is the Einstein universe. We first list the Killing vector
fields which are also Noether symmetries
\begin{align*}
&
        \mathbf{X_{4,1}}=r\frac{\partial}{\partial
t}+t\frac{\partial}{\partial
r},\quad \mathbf{X_{5,1}}=\frac{\partial}{\partial r},\\
& \mathbf{X_{4,2}}=-\beta re^{-t/\beta}\frac{\partial}{\partial
t}+r^2e^{-t/\beta}\frac{\partial}{\partial r},\quad
\\&
\mathbf{X_{5,2}}=\beta re^{t/\beta}\frac{\partial}{\partial
t}+r^2e^{t/\beta}\frac{\partial}{\partial r}\\&
\mathbf{X_{4,3}}=\frac{b}{b+r}e^{\frac{-t}{b}}\frac{\partial}{\partial
t}+e^{\frac{-t}{b}}\frac{\partial}{\partial r},\quad
\mathbf{X_{5,3}}=-\frac{b}{b+r}e^{(t/b)}\frac{\partial}{\partial
t}+e^{(t/b)}\frac{\partial}{\partial
r},\\
&
\mathbf{X_{4,4}}=\sqrt{b^2-r^2}\sin\phi\sin\theta\frac{\partial}{\partial
r}-
\frac{\sqrt{b^2-r^2}}{r}\cos\theta\sin\phi\frac{\partial}{\partial
\theta}+\frac{\sqrt{b^2-r^2}}{r\sin\theta}\cos\phi\frac{\partial}{\partial
\phi},
\\
&
\mathbf{X_{5,4}}=\sqrt{b^2-r^2}\cos\phi\sin\theta\frac{\partial}{\partial
r}-
\frac{\sqrt{b^2-r^2}}{r}\cos\theta\cos\phi\frac{\partial}{\partial
\theta}-\frac{\sqrt{b^2-r^2}}{r\sin\theta}\sin\phi\frac{\partial}{\partial
\phi},
\\
& \mathbf{X_{6,4}}=\sqrt{b^2-r^2}\cos\theta\frac{\partial}{\partial
r}- \frac{\sqrt{b^2-r^2}}{r}\sin\theta\frac{\partial}{\partial
\theta},
\end{align*}
 and the Noether symmetries corresponding to non-trivial guages are
\begin{align*}
&
\mathbf{Y_{1,1}}=s\frac{\partial}{\partial
t}, \quad A_{1,1}=2t\\
&
 \mathbf{Y_{2,1}}=s\frac{\partial}{\partial r}, \quad A_{2,1}=2r\\
&
\mathbf{Y_{1,2}}=-\frac{rse^{-t/\beta}}{\beta^3}\frac{\partial}{\partial
t}+\frac{r^2se^{-t/\beta}}{\beta^4}\frac{\partial}{\partial r}, \quad A_{1,2}=\frac{2e^{-t/\beta}}{r}\\
&
  \mathbf{Y_{2,2}}=\frac{rse^{t/\beta}}{\beta^3}\frac{\partial}{\partial
t}+\frac{r^2se^{t/\beta}}{\beta^4}\frac{\partial}{\partial r}, \quad A_{1,2}=\frac{2e^{t/\beta}}{r}\\
&
\mathbf{Y_{1,3}}=-\frac{bs}{2(b+r)}e^{(-t/b)}\frac{\partial}{\partial
t}-\frac{s}{2}e^{(-t/b)}\frac{\partial}{\partial r}, \quad A_{1,3}=(b+r)e^{(-t/b)}\\
&
\mathbf{Y_{2,3}}=\frac{bs}{2(b+r)}e^{(t/b)}\frac{\partial}{\partial
t}-\frac{s}{2}e^{(t/b)}\frac{\partial}{\partial
r}, \quad A_{2,3}=(b+r)e^{(t/b)}\\
& \mathbf{Y_{1,4}}=s\frac{\partial}{\partial t}, \quad  \quad A=2t.
\end{align*}The Lie algebra of symmetries of metric (2) above is
\begin{align*}&[\mathbf{X_1},\mathbf{X_3}]=\mathbf{X_2},\quad  [\mathbf{X_2},\mathbf{X_3}]=-\mathbf{X_1}
 ,\quad,[\mathbf{X_0},\mathbf{X_{4,2}}]=\frac{1}{\alpha}\mathbf{X_{4,2}},\quad [\mathbf{X_{4,2}},\mathbf{X_{5,2}}]=-\mathbf{X_{4,2}}\\&
[\mathbf{X_0},\mathbf{X_{5,2}}]=\frac{1}{\alpha}\mathbf{X_{5,2}},\quad
[\mathbf{X_0},\mathbf{Y_{1,2}}]=-\frac{1}{\alpha}\mathbf{Y_{1,2}},\quad
[\mathbf{X_0},\mathbf{Y_{2,2}}]=\frac{1}{\alpha}\mathbf{Y_{2,2}},\\&
[\mathbf{Y_0},\mathbf{Y_{1,2}}]=\frac{1}{\alpha^4}\mathbf{X_{4,2}},\quad
[\mathbf{Y_0},\mathbf{Y_{2,2}}]=\frac{1}{\alpha^4}\mathbf{X_{5,2}},\quad
 [\mathbf{X_i},\mathbf{X_j}]=0,\quad [\mathbf{X_i},\mathbf{Y_0}]=0,\\& [\mathbf{Y_i},\mathbf{Y_j}]=0,\qquad otherwise\end{align*}
\begin{table}[H]
\begin{center}
\captionof{table}{First Integrals of Spacetimes admitting 9 Noether
Symmetries}
\begin{tabular}{|c|c|}\hline

                Gen & First integrals\\
                \hline
$\mathbf{X_{4,1}}$,\quad$\mathbf{X_{5,1}}$&$\phi_5=t\dot{r}-r\dot{t}$,\qquad$\phi_6=\dot{r}$ \\
\hline
$\mathbf{Y_{1,1}}$,\quad$\mathbf{Y_{2,1}}$&$\phi_7=2[t-s\dot{t}]$,\qquad $\phi_8=r-s\dot{r}$\\
\hline $\mathbf{X_{4,2}}$,\quad
$\mathbf{X_{5,2}}$&$\phi_5=2e^{-t/\alpha}\alpha^3[\frac{\dot{t}}{r}+\frac{\alpha\dot{r}}{r^2}]$,
\qquad$\phi_6=2e^{t/\alpha}\alpha^3[-\frac{\dot{t}}{r}+\frac{\alpha\dot{r}}{r^2}]$\\
\hline
$\mathbf{Y_{1,2}}$,\quad$\mathbf{Y_{2,2}}$&$\phi_7=2se^{-t/\alpha}[\frac{\dot{t}}{r\alpha}+\frac{\dot{r}}{r^2}]+\frac{2e^{-t/\alpha}}{r}$,\qquad
$\phi_8=2se^{t/\alpha}[\frac{-\dot{t}}{\alpha r}+\frac{\dot{r}}{r^2}]+\frac{2e^{t}}{r}$ \\
\hline
$\mathbf{X_{4,3}}$,\quad$\mathbf{X_{5,3}}$&$\phi_5=e^{-t/b}\left(-\frac{\dot{t}(b+r)}{b}+\dot{r}\right)$,
\qquad$\phi_6=e^{t/b}\left(\frac{\dot{t}(b+r)}{b}+\dot{r}\right)$\\
\hline $\mathbf{Y_{1,3}}$,\quad $\mathbf{Y_{2,3}}$
&$\phi_7=e^{-t/b}\left(\frac{s\dot{t}(b+r)}{b}+s\dot{r}+(b+r)\right)$,\qquad $\phi_8=e^{t/b}\left(\frac{s\dot{t}(b+r)}{b}-s\dot{r}+(b+r)\right)$\\
\hline $\mathbf{X_{4,4}}$ &
$\phi_5=\frac{b^2\dot{r}\sin\phi\sin\theta}{\sqrt{b^2-r^2}}-
r\dot{\theta}\sqrt{b^2-r^2}\cos\theta\sin\phi+r\dot{\phi}\sqrt{b^2-r^2}\sin\theta\cos\phi$\\
\hline
$\mathbf{X_{5,4}}$&$\phi_6=\frac{b^2\dot{r}\cos\phi\sin\theta}{\sqrt{b^2-r^2}}-
r\dot{\theta}\sqrt{b^2-r^2}\cos\theta\cos\phi+r\dot{\phi}\sqrt{b^2-r^2}\sin\theta\sin\phi$\\
\hline
$\mathbf{X_{6,4}}$,\quad$\mathbf{Y_{1,4}}$&$\phi_7=\frac{b^2\dot{r}\cos\theta}{\sqrt{b^2-r^2}}-
r\dot{\theta}\sqrt{b^2-r^2}\sin\theta$,\qquad$\phi_8=t-s\dot{t}$\\
\hline
\end{tabular}\end{center} \end{table}
\textbf{V- Spherically Symmetric Static Spacetimes with Eleven
Noether Symmetries}
\newline
The famous de-Sitter metric turns out to be the only case with eleven Noether symmetries. Except $\mathbf{Y_0}$ all others are the Killing vctors.
\begin{equation}
1: \hspace{1cm}\mbox{d}s^2=\left (1-\frac{r^2}{b^2}\right )\mbox{d}t^2-\frac{\mbox{d}r^2}{\left (1-\frac{r^2}{b^2} \right)}-r^2\mbox{\mbox{d}}\Omega^{2}.\label{50}
\end{equation}
We have the following Noether symmetries along with the minimal set
of Noether symmetries for the metric given by $(\ref{50})$
\begin{align*}
\mathbf{X_4} =&
\frac{br\sin\phi\sin\theta\cos(t/b)}{\sqrt{b^2-r^2}}\frac{\partial}{\partial t}+
\sin(t/b)\sqrt{b^2-r^2}\left (\sin\theta\sin\phi\frac{\partial}{\partial
r}+{r}\cos\theta\sin\phi\frac{\partial}{\partial
\theta}+
\frac{\cos\phi}{r\sin\theta}\frac{\partial}{\partial
\phi}\right ),
\\
\mathbf{X_5}=&
\frac{br\cos\phi\sin\theta\cos(t/b)}{\sqrt{b^2-r^2}}\frac{\partial}{\partial t}+
\sin(t/b)\sqrt{b^2-r^2}\left ( \sin\theta\cos\phi\frac{\partial}{\partial
r}+\frac{\cos\theta\cos\phi}{r}\frac{\partial}{\partial
\theta}-\frac{\sin\phi}{r\sin\theta}\frac{\partial}{\partial
\phi}\right ),
\\
\mathbf{X_6}=&\frac{-br\sin\phi\sin\theta\sin(t/b)}{\sqrt{b^2-r^2}}\frac{\partial}{\partial t}+
\cos(t/b)\sqrt{b^2-r^2}\left (\sin\theta\sin\phi\frac{\partial}{\partial
r}+{r}\cos\theta\sin\phi\frac{\partial}{\partial
\theta}+
\frac{\cos\phi}{r\sin\theta}\frac{\partial}{\partial
\phi}\right ),
\\
\mathbf{X_7}=&\frac{-br\cos\phi\sin\theta\sin(t/b)}{\sqrt{b^2-r^2}}\frac{\partial}{\partial t}+
\cos(t/b)\sqrt{b^2-r^2}\left ( \sin\theta\cos\phi\frac{\partial}{\partial
r}+\frac{\cos\theta\cos\phi}{r}\frac{\partial}{\partial
\theta}-\frac{\sin\phi}{r\sin\theta}\frac{\partial}{\partial
\phi}\right ),
\\
\mathbf{X_8}=&
\frac{br\cos\theta\cos(t/b)}{\sqrt{b^2-r^2}}\frac{\partial}{\partial t}+
\sin(t/b)\sqrt{b^2-r^2}\left (\cos\theta\frac{\partial}{\partial
r}-\frac{\sin\theta}{r}\frac{\partial}{\partial
\theta}\right ),
\\
\mathbf{X_9}=&
\frac{-br\cos\theta\sin(t/b)}{\sqrt{b^2-r^2}}\frac{\partial}{\partial t}+
\cos(t/b)\sqrt{b^2-r^2}\left (\cos\theta\frac{\partial}{\partial
r}-\frac{\sin\theta}{r}\frac{\partial}{\partial
\theta}\right ).
\end{align*}
\\ The first integrals
are given in the following
Table 6.
\begin{table}[H]
\captionof{table}{First Integrals}
\begin{center}\small\begin{tabular}{|c|c|}\hline

                Gen & First integrals \\
                \hline
$\mathbf{X_5}$&$
\phi_5=-\frac{r}{b}\sqrt{b^2-r^2}\sin\phi\sin\theta\cos(t/b)\dot{t} + \frac{b^2}{\sqrt{b^2-r^2}}\sin\phi\sin\theta\sin(t/b)\dot{r}+
$\\
& $
r\sqrt{b^2-r^2}\cos\theta\sin\phi\sin(t/b)\dot{\theta} +r\sqrt{b^2-r^2}\sin\theta\cos\phi\sin(t/b)\dot{\phi}$
\\
\hline
$\mathbf{X_6}$&$\phi_6 = -\frac{r}{b}\sqrt{b^2-r^2}\cos\phi\sin\theta\cos(t/b)\dot{t} + \frac{b^2}{\sqrt{b^2-r^2}}\cos\phi\sin\theta\sin(t/b)\dot{r}
+$\\
&
$r\sqrt{b^2-r^2}\cos\theta\cos\phi\sin(t/b)\dot{\theta}-r\sqrt{b^2-r^2}\sin\theta\sin\phi\sin(t/b)\dot{\phi}$\\
\hline
$\mathbf{X_7}$ & $\phi_8=\frac{r}{b}\sqrt{b^2-r^2}\sin\phi\sin\theta\sin(t/b)\dot{t} + \frac{b^2}{\sqrt{b^2-r^2}} \sin\phi\sin\theta\cos(t/b)\dot{r}+
$\\
& $ r\sqrt{b^2-r^2}\cos\theta\sin\phi\cos(t/b)\dot{\theta}+r\sqrt{b^2-r^2}\sin\theta\cos\phi\cos(t/b)\dot{\phi}$\\
\hline
$\mathbf{X_8}$&$\phi_9=-\frac{r}{b}\sqrt{b^2-r^2}\cos\phi\sin\theta\cos(t/b)\dot{t} + \frac{b^2}{{\sqrt{b^2-r^2}}}\cos\phi\sin\theta\sin(t/b)\dot{r}+
$
\\
&
$r\sqrt{b^2-r^2}\cos\theta\cos\phi\sin(t/b)\dot{\theta}-r\sqrt{b^2-r^2}\sin\theta\sin\phi\sin(t/b)\dot{\phi}$
\\
\hline
$\mathbf{X_9}$&$\phi_7 = -\frac{r}{b}\sqrt{b^2-r^2}\cos\theta\cos(t/b)\dot{t}+\frac{b^2}{\sqrt{b^2-r^2}}\cos\theta\sin(t/b)\dot{r}
-r\sqrt{b^2-r^2}\sin\theta\sin(t/b)\dot{\theta}$\\
\hline
$\mathbf{X_{10}}$&$\phi_{10}=\frac{r}{b}\sqrt{b^2-r^2}\cos\theta\sin(t/b)\dot{t}+\frac{b^2}{{\sqrt{b^2-r^2}}}\cos\theta\cos(t/b)\dot{r}
-r\sqrt{b^2-r^2}\sin\theta\cos(t/b)\dot{\theta}$\\
\hline
\end{tabular}\end{center}
\end{table}

\textbf{V - Spherically Symmetric Static Spacetimes with Seventeen Noether Symmetries}
\newline
It is the famous Minkowski metric that represents a flat spacetime and admits seventeen Noether symmetries. The list of all symmetries is given in \cite{14}, while the first integrals are mentioned in \cite{4} in Cartesian coordinate system.
\section{Conclusion}
In this paper a complete list of classification of spherically
symmetric static spacetimes is given. It is seen that spherically
symmetric static spacetimes may have $5, 6, 7, 9, 11,$ or $17$
Noether symmetries. A few examples of spacetimes having minimal
(i.e. 5) Noether symmetries are given in Table 1. Briefly, there
appear two cases in which spacetimes admit six, three cases having
seven, five cases admitting nine (including the Bertotti-Robinson
and the Einstein metrics) whereas there appear only one case of
eleven (which is the famous de-Sitter spacetime) and seventeen
(Minkowaski spacetime) Noether symmetries. Just like plane symmetric
static spacetimes, for spherically symmetric static spacetimes the
minimum number of Noether symmetries are five and the maximum number
of Noether symmetries are seventeen, while the minimum number of
isometries are four and the maximum number of isometries are ten. The work on classification problem of
cylindrically symmetric static spacetimes with
respect to their Noether symmetries is in progress.

There are three new cases that we have not seen in the literature,
it is necessary to mention them briefly. We also list the non-zero components of Riemannian tensors, Ricci
tensors and Ricci scalar.\\
The first case is
\begin{equation}
(i):\qquad ds^2=\sec^2\left (\frac{r}{a}\right)
(dt^2-dr^2)-\mbox{\mbox{d}}\Omega^{2}
\end{equation}
in which spacetime has seven Noether symmetries and six Killing
vectors.\\ The Ricci scalar, Ricci tensors and
 components of Riemann tensors are
\begin{align*}
&
R_{\mbox{scalar}}=\frac{2(a^2-1)}{a^2},\\
&R_{00}=-\frac{\sec^2\frac{r}{a}}{a^2},\,
R_{11}=\frac{\sec^2\frac{r}{a}}{a^2},\, R_{22}=-1,
R_{33}=-\sin^2\theta,\\
&  R_{0101}=-\frac{\sec^4\frac{r}{a}}{a^2},\,
R_{2323}=-\sin^2\theta.
\end{align*}
In second case, spacetime
\begin{equation}
(ii):\qquad
ds^2=\left(\frac{a}{r}\right)^2(dt^2-dr^2)-\mbox{\mbox{d}}\Omega^{2}
\end{equation}
also has seven Noether symmetries and six Killing
Vectors. It may be pointed that the above metric takes the form
\begin{align*}
&
r^2ds^2= a^{2}(dt^2-dr^2)-r^2\mbox{\mbox{d}}\Omega^{2},\\
& ds^2= \frac{1}{r^2}d\widetilde{s},\,\,
\end{align*}
where $d\widetilde{s}$ is the standard Minkowski metric, therefore,
$(21)$ represents a conformal Minkowski metric with conformal factor
$1/r^2$.\\ The Ricci scalar, Ricci tensors and Riemann tensors are
given below
\begin{align*}
&
R_{\mbox{scalar}}=\frac{(a^2-1)}{a^2},\\
& R_{00}=-\frac{1}{r^2},\,\quad R_{11}=\frac{1}{r^2},\, R_{22}=-1,
R_{33}=-\sin^2\theta,\\
& R_{0101}=-\frac{a^2}{r^4},\quad R_{2323}=-\sin^2\theta.
\end{align*}
The third case is
\begin{equation}
(iii):\qquad
ds^2=\left(\frac{a}{r}\right)^2dt^2-\left(\frac{a}{r}\right)^4dr^2-\mbox{\mbox{d}}\Omega^{2}
\end{equation}
which has nine Noether symmetries.\\ The corresponding Ricci scalar,
Ricci tensors and Riemann tensors are the following
\begin{align*}
&
R_{\mbox{scalar}}=2,\\
& R_{22}=-1,\quad R_{33}=-\sin^2\theta,
\\&
R_{2323}=-\sin^2\theta.
\end{align*}

\end{document}